\documentclass[11pt]{article} % use larger type; default would be 10pt

\usepackage[utf8]{inputenc} % set input encoding (not needed with XeLaTeX)

\usepackage{geometry} % to change the page dimensions
\usepackage{enumerate}
\usepackage{amsmath}
\usepackage{amssymb}
\usepackage{dsfont}
\usepackage{verbatim}
\usepackage{graphicx}
\usepackage{placeins}
\usepackage{hyperref}

\numberwithin{equation}{section}
\numberwithin{figure}{section}
\numberwithin{table}{section}

\DeclareMathOperator*{\argmin}{\arg\!\min}

\title{A Flexible Commodity Skew Model with Maturity Effects} 
\author{\Large Orcan \"Ogetbil\footnote{orcan.ogetbil@wellsfargo.com} \ 
and Bernhard Hientzsch\footnote{bernhard.hientzsch@wellsfargo.com}\\
{\small Corporate Model Risk, Wells Fargo Bank}}

\date{} % to remove the date from the cover

\begin{document}

\maketitle
\setcounter{secnumdepth}{3}
\begin{abstract}

We propose a non-parametric extension with leverage functions to the Andersen
commodity curve model. We calibrate this model to market data for WTI and NG
including option skew at the standard maturities. While the model can be
calibrated by an analytical formula for the deterministic rate case, the
stochastic rate case demands estimation of an expectation for which we employ
Monte Carlo simulation. We find that the market smile is captured for the
deterministic rate case; and with relatively low number of paths, for the
stochastic rate case. Since there is typically at most one standard maturity
with liquid volatility data for each futures contract, there is flexibility on
the shape of nonstandard maturity implied volatility and how the total implied
variance accumulates. We equip the model with different total implied variance
accumulators to demonstrate that flexibility.

\end{abstract}

\section{Introduction}

For both pricing and traded risk management such as CVA for instruments based 
on futures of different maturities, one needs commodity curve models with various 
properties.  In our particular setting, we are looking for a model that starts 
from a model in which a small number of Brownian factors drives 
the entire curve and captures ATM volatility and (imperfect) correlation (like the Andersen
model) but extends it to a model in which a given implied volatility slice or a number of calibration 
European options on the standard expiry for each futures contract is repriced as 
well as possible for each futures contract separately from each other,
regardless whether they follow some parametric form or not, as long as they are
arbitrage free.
We would like to select a model which we can calibrate by either an explicit formula 
from given volatility market data or by something that can be computed by Monte-Carlo
through computation of expectations or through Monte-Carlo based regressions, rather 
than by approaches involving more computationally or implementationally complex
and expensive methods such as particle filters, optimization through PDE/PIDE or FFT/Integral 
solvers, or similar methods. We also prefer models that can be easily evaluated and 
simulated at each needed time (as needed for exposure computations) with standard 
methodologies. 

Since for each futures contract, there is at most one liquidly traded option
maturity\footnote{For 
some delivery months, liquid option trading for the 
standard maturity might not have started yet and hence those contracts 
might not have liquid volatility data yet. In those circumstances, we use 
an accumulator derived from the implied volatility of 
the next liquidly traded longer contract.}, 
we want to have the freedom to accumulate the total
implied variance from that maturity through time in a flexible  way which would
allow the model to mimick stylized facts in volatility behavior in the
observational and/or risk-neutral measure. (Should there be more than one
liquidly trade option maturity or should there be internally marked volatility
slices at different maturities, we similarly could accumulate the total implied
variance through time but now constrained by data on more slices.)  We also want
to allow additional stochastic variance or volatility factors or processes if
desired.

Without skew modeling, the movement of commodity futures was modeled with two Brownian motions
previously in \cite{Gabillon1991}, where the futures prices are derived from the
modeled but not directly observable quantities such as spot commodity and a
very long term futures contract that represents the long maturity asymptotic
behavior.

More recently Andersen \cite{Andersen2010} suggested three different approaches
to add skew to the model -- jump diffusion models, stochastic volatility models, and regime-switch models (possibly combined);
and further discussing regime-switch models, all on the level of the factors driving the curve 
and not associated to futures of particular maturities. Jump diffusion models lead to Partial Integral 
and Differential Equations (PIDE) and stochastic volatility models lead to FFT and integral equation 
such as in the famous Heston model. PIDE are quite involved to treat well, stochastic volatility models treated by
FFT/characteristic function methods also require careful treatement. Stochastic volatility models 
typically cannot fit skews everywhere well enough; one typically needs local stochastic volatility 
models to fit skews well. Regime-switch model lead to a mixture of Black-Schole/lognormal formulas so they are 
easy to compute but mixtures of lognormals cannot fit all skews well enough. All of these 
models modify the form of the entire futures curve, so one has to fit skews
across all maturities, either at the same time or one after the other. 
Andersen \cite{Andersen2010} does not give details about calibration or pricing with 
jump diffusion or stochastic volatility extensions. He does show an example fitting for 
his suggested regime-switch model. However, he did not have reliable option skew data 
from his trading desk and the fit to the option skew data he used was only approximate. 

Following \cite{Gabillon1991}, modeling a fictitious spot commodity process has
been studied at several occasions in the literature. In \cite{NPS2020} a
one-factor fictitious spot model is fit to the given volatility slices at the
standard maturities. The authors need to introduce some freedom in the drift of
the spot to be able to calibrate to the market data.
Since it is a one-factor model, different futures will be perfectly correlated,
but the volatility of the fictitious spot can be computed by an extended Dupire
equation. They quickly mention extensions to several factors (in which the
one-factor volatility term is redistributed to several factors), and to
stochastic volatility factors in addition to the leverage function, but do not
give details or results. There is no freedom to shape or redistribute total
implied variance (or to model any non-front contract) and there is perfect
correlation.
\cite{MNPV2022} introduces a two-factor stochastic local volatility fictitious
spot model (two spot processes) so that futures options and index options can be
priced, modeling the front two contracts at different times, capturing the
correlation between them, and calibrating the models with particles filters.
While the front two contracts can be modeled, no other futures contracts are
modeled in detail.

\cite{schneider2018seasonal,schneider2015seasonal,schneider2014samuelson} introduce 
stochastic volatility multi-factor models where each Brownian curve driving factor 
is associated with its own CIR-type stochastic volatility factor with a time-dependent
seasonal mean. They introduce the model both in risk-neutral and observational 
measure, describe FFT based pricing (and calibration) for the risk-neutral setting, 
and a Kalman filter for its use under the observational measure. They explicitly 
introduce terms to introduce maturity ("Samuelson") effects. 
They demonstrate reasonable good fitting of option skews but not all option quotes can be fit 
(see \cite[version 3: Figure 3 on page 28]{schneider2014samuelson}). 
Pricing and calibration in these models require FFT and/or direct integration 
approaches with careful treatment of characteristic functions and domain or contour 
of integration and possibly need damping. We are looking for a model set-up which 
allows calibration and pricing with simpler methodologies.   

A model in both physical $\mathbb{P}$ and risk-neutral $\mathbb{Q}$ measures is
presented in \cite{LB2021} that can at most fit at-the-money (ATM) volatility
and has no more degrees of freedom for the skew.  However, this model could be
extended to capture skew similarly how we add skew to the Andersen model in this
paper.
\cite{BBV2013} models the skew in the observational measure and discusses option
pricing. It involves very heavy mathematical machinery and does not fit easily
in the standard simulation set-ups. It does not demonstrate calibration to
option prices.
Futures curves driven by time-changed Ornstein-Uhlenbeck processes are modeled
in \cite{LL2012}, including with additional stochastic volatility which can also
fit common skew shapes in their examples. Modeling and calibration are quite
involved and time-changed processes do not fit well into the standard simulation
set-up.

Instead, in our work we add separate leverage functions as multipliers for each
modeled futures, and discuss how to calibrate the resulting model by formula
(for deterministic rates) or by computing MC expectations (for stochastic
rates); and will review how this model satisfies all of our requirements as
discussed earlier.
 
The structure of the paper is as follows. Section \ref{sec:Andersen_review}
reviews Andersen's two-factor model and shows how it calibrates to recent
market data for WTI and NG. In Section \ref{sec:Andersen_LV}, we extend this
model with leverage functions. We outline steps for calibration of leverage
functions for both deterministic and stochastic rate cases, and demonstrate that
the calibrated model captures the market smile. We propose a method to 
shape the forward implied volatility (through the remaining total implied 
variance) of options on futures with longer deliveries according to the 
implied volatility of corresponding options on futures with shorter deliveries 
and also mention other ways to control the accumulation of total implied 
variance for futures options with longer deliveries.  
In Section \ref{sec:discussion} we briefly discuss our findings.

% \tableofcontents
\section{Andersen's Markov Diffusion Model on Commodity
Futures}\label{sec:Andersen_review}
%\newpage{}

We start with reviewing the diffusion model proposed by Andersen \cite{Andersen2010}.
Motivated by principal component analysis on historical futures returns and
practicality reasons, futures price curve movements are governed by two Brownian
motions. Let $W_1$ and $W_2$ be \emph{independent} Brownian motions under
risk-neutral measure $\mathbb{Q}$ of filtered probability space $\left(\Omega,
\mathcal{F}, \left\{\mathcal{F}_t, t \geq 0 \right\}, \mathbb{Q}\right)$.
Let us denote the futures delivery times by $T_j, j=1,\ldots,J$. Here $J$
depends on data availability and is different for each commodity underlier. The
price of a futures contract $F_j(t)$ with delivery time $T_j$ evolves by
\begin{equation}
dF_j(t) = F_j(t) \left[\sigma_1(t, T_j) dW_1(t) + \sigma_2(t, T_j)
dW_2(t)\right].\label{eqn:andersen_sde}
\end{equation}
Here the deterministic functions $\sigma_1(t, T)$ and $\sigma_2(t, T)$ are
given by
\begin{equation}
\sigma_1(t, T) = e^{b(T)-\kappa(T-t)} h_1 + e^{a(T)} h_\infty,\ \ \ \ 
\sigma_2(t, T) = e^{b(T)-\kappa(T-t)} h_2,
\end{equation}
where the model parameters $h_1, h_2, h_\infty$ are constants; $\kappa \geq 0$
is mean reversion speed; $a(T), b(T)$ are seasonality adjustment functions
oscillating around zero at an annual frequency. The following parametrization
allows us to write the model parameters in terms of more intuitive quantities.
\begin{equation}
\sigma_0 \equiv \sqrt{(h_1 + h_\infty)^2 +h_2^2},\ \ \ \sigma_\infty \equiv h_\infty,\ \ \ 
\rho_\infty \equiv \frac{h_1 +h_\infty}{\sqrt{(h_1 + h_\infty)^2 +h_2^2}}.
\end{equation}
In this parametrization, $\sigma_0$ and $\sigma_\infty$ represent the short term
and long term futures price volatilities in the absence of seasonality, and
$\rho_\infty$ is the correlation between the price changes of the spot and the
long-end of the futures curve.
The at-the-money market implied volatility is related to the model parameters as
\cite{Andersen2010},
\begin{align}
\Sigma^{\text{ATM}}(T)^2 = (h_1^2 + h_2^2) e^{2b(T)}\frac{1 -
e^{-2\kappa T}}{2\kappa T} + 2 h_{\infty} h_1 e^{a(T)+b(T)}\frac{1 - e^{-\kappa
T}}{\kappa T} + h^2_{\infty} e^{2a(T)}.\label{eq:sigmaatm}
\end{align}

\begin{figure}[ht!]
    \centering
    \includegraphics[width=0.495\textwidth]{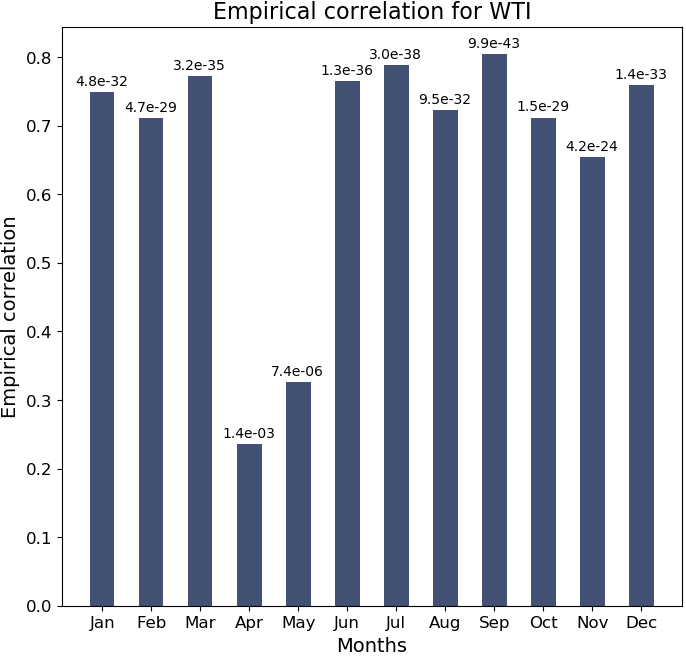}
    \includegraphics[width=0.495\textwidth]{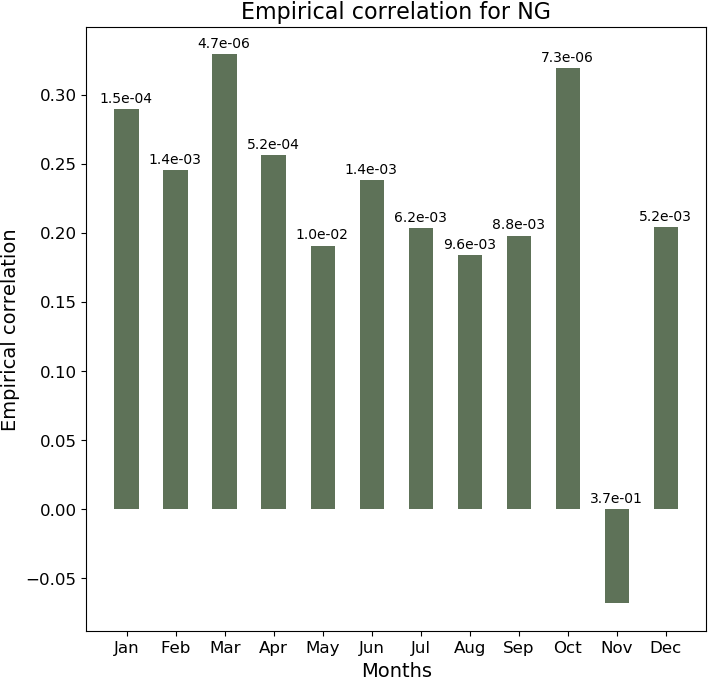}
    \caption{Estimating $f_\infty (t)$ empricially for WTI and NG data between
    2013-01-01 and 2021-12-31, as a function of the calendar month of $t$. The
    number above each bar is the corresponding p-value.}
    \label{fig:rhoinf_months}
\end{figure}

\begin{figure}[ht!]
    \centering
    \includegraphics[width=\textwidth]{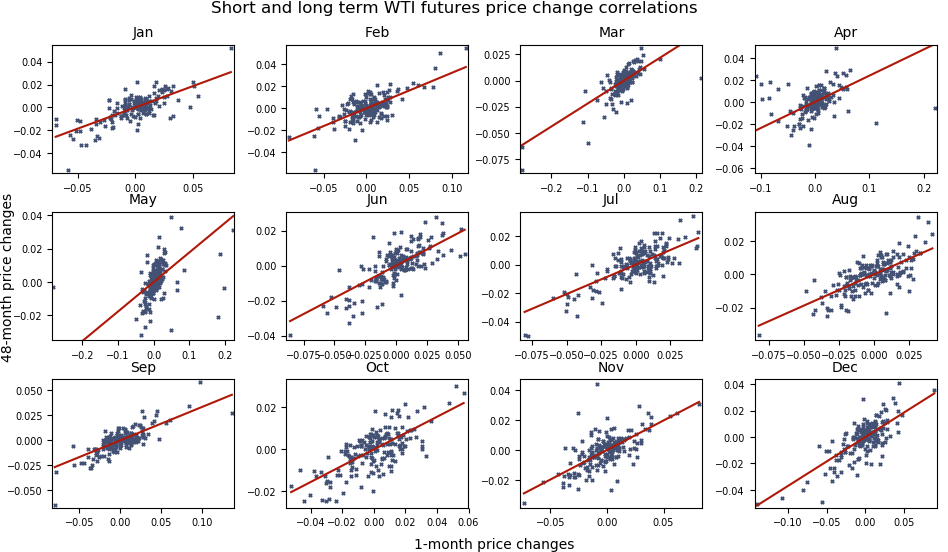}
    \newline\vspace{5pt}
    \includegraphics[width=\textwidth]{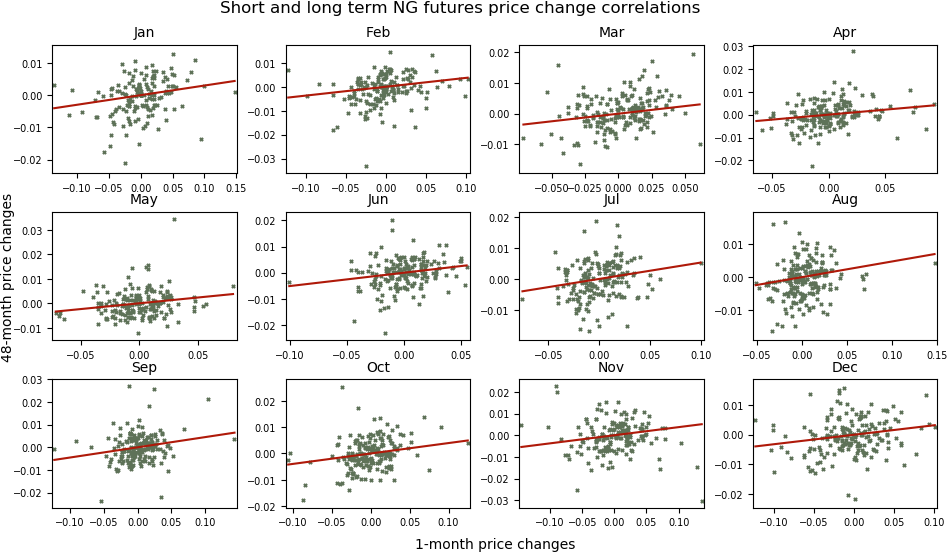}
    \caption{Daily price changes of 1-month and 48-month expiry futures of WTI
    and NG data between 2013-01-01 and 2021-12-31, for each month of the year.
    The red lines correspond to the estimated values $\rho_\infty = 0.7$ for WTI
    and $\rho_\infty = 0.2$ for NG.}
    \label{fig:corrs_months}
\end{figure}

\begin{figure}[ht!]
    \centering
    \includegraphics[width=0.49\textwidth]{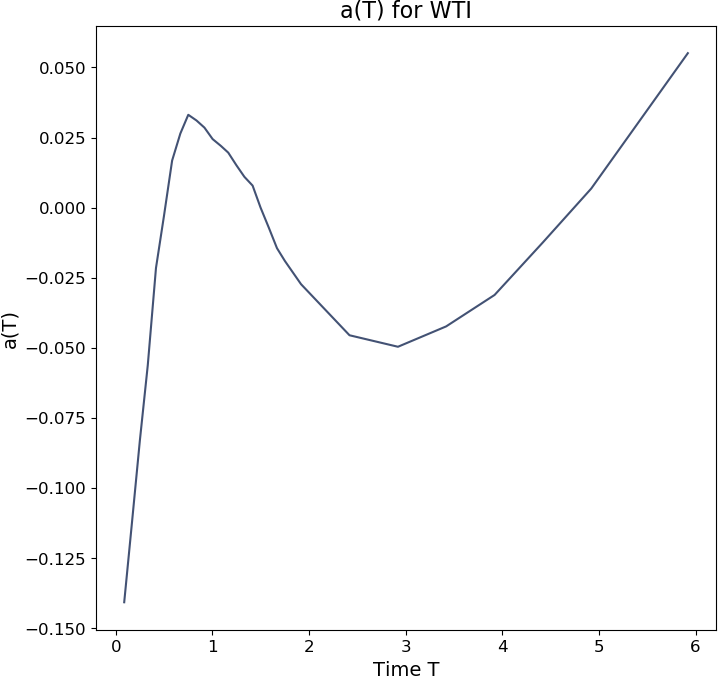}
    \includegraphics[width=0.49\textwidth]{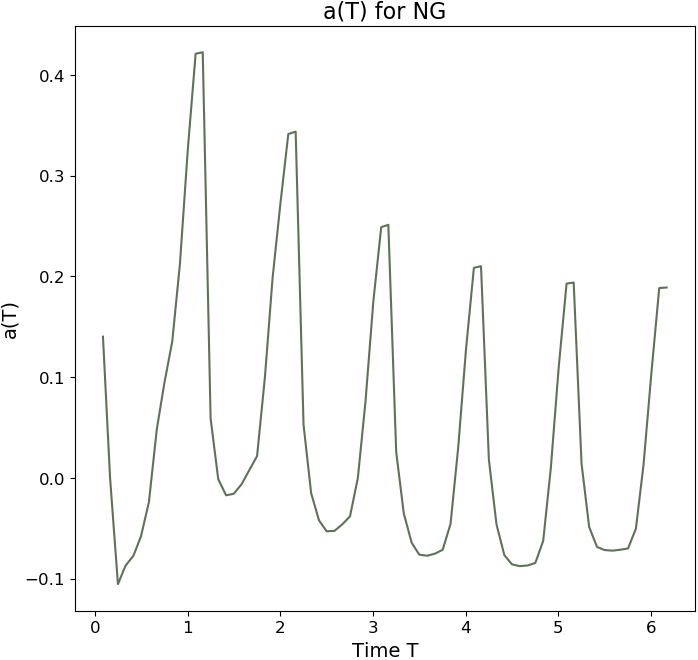}
    \caption{Seasonality adjustment function $a(T)$ for WTI and NG computed
    by Lemma 10 of \cite{Andersen2010} based on implied volatility data as of
    2021-12-31.}
    \label{fig:a_T}
\end{figure}

We calibrate the model parameters to the market data following the recipe
outlined in \cite{Andersen2010}. First, we study the correlation structure.
Setting $X(t, T_j) \equiv \log F_j(t)$, through the correlation function
\begin{equation}
\rho(t, \Delta_1, \Delta_2) = \text{corr}\left(dX(t, t + \Delta_1), dX(t, t +
\Delta_2)\right)
\end{equation}
we define a function to measure the time dependence of correlations
\begin{equation}
f_\infty (t) \equiv \lim_{\Delta \rightarrow \infty} \rho(t,
0, \Delta).
\end{equation}
We analyze futures returns data for WTI and NG to estimate $f_\infty (t)$
empirically using $\rho(t, \text{1 month}, \text{48 months})$ as a proxy to
this quantity. According to Figure \ref{fig:rhoinf_months}, the correlation
structure does not show a clear seasonal trend, thus we can model it as
time-stationary, $f_\infty (t) = \rho_\infty$, and $a(T) = b(T)$. We estimate
$\rho_\infty = 0.7$ for WTI and $\rho_\infty = 0.2$ for NG. In Figure
\ref{fig:corrs_months}, we see that these estimations fit the price change data
reasonably well, even on those months with lower empirical correlations.
% Script used:
% scripts/WTI_calibrate_Andersen.ipynb
% scripts/NG_calibrate_Andersen.ipynb
\FloatBarrier
The rest of the parameters are calibrated to the market implied volatility data
as of 2021-12-31 by \eqref{eq:sigmaatm} as $\kappa=0.2657, h_1 = 0.2365, h_2 = 
0.2970, h_\infty = 0.0546$ for WTI and $\kappa=2.3562, h_1 = -0.0634, h_2 =
0.5854, h_\infty = 0.1829$ for NG, with $a(T)$ as given in Figure
\ref{fig:a_T}.
% Script used:
% scripts/WTI_calibrate_Andersen.ipynb
% scripts/NG_calibrate_Andersen.ipynb

\FloatBarrier

\begin{figure}[ht!]
    \centering
    \includegraphics[width=0.495\textwidth]{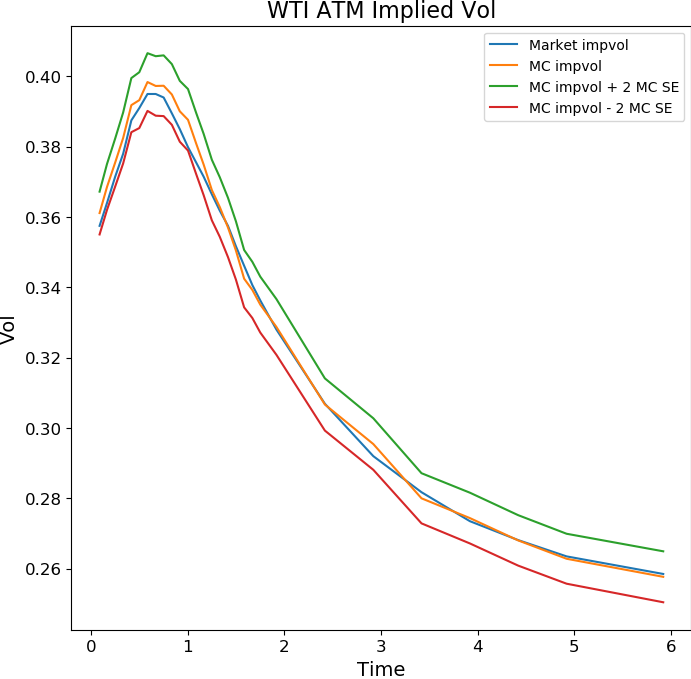}
    \includegraphics[width=0.495\textwidth]{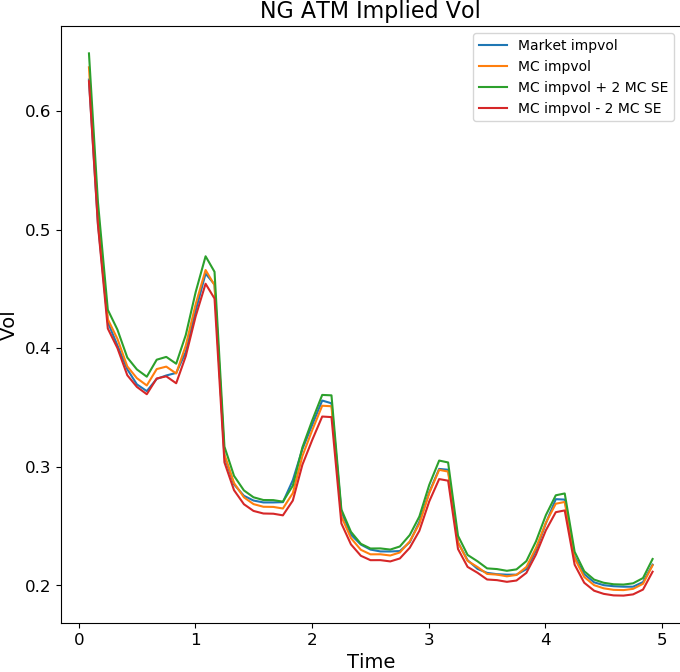}
    \caption{Market and Monte Carlo at-the-money (ATM) implied volatility
    for WTI and NG}
    \label{fig:atm_vols}
\end{figure}

We simulate the model with the above calibrated parameters to price vanilla call
options. First we price at-the-money vanilla call options at various expiries,
and by inverting the Black formula we evaluate the implied volatility from Monte
Carlo prices as well as Monte Carlo prices bumped by $\pm 2$ Monte Carlo
standard errors.
Figure \ref{fig:atm_vols} shows that the ATM market implied volatility is well
recovered by the model with the calibrated parameters.
% Script used:
% scripts/WTI_simulate_Andersen.ipynb scripts/NG_simulate_Andersen.ipynb

\FloatBarrier

\begin{figure}[ht!]
    \centering
    \includegraphics[width=\textwidth]{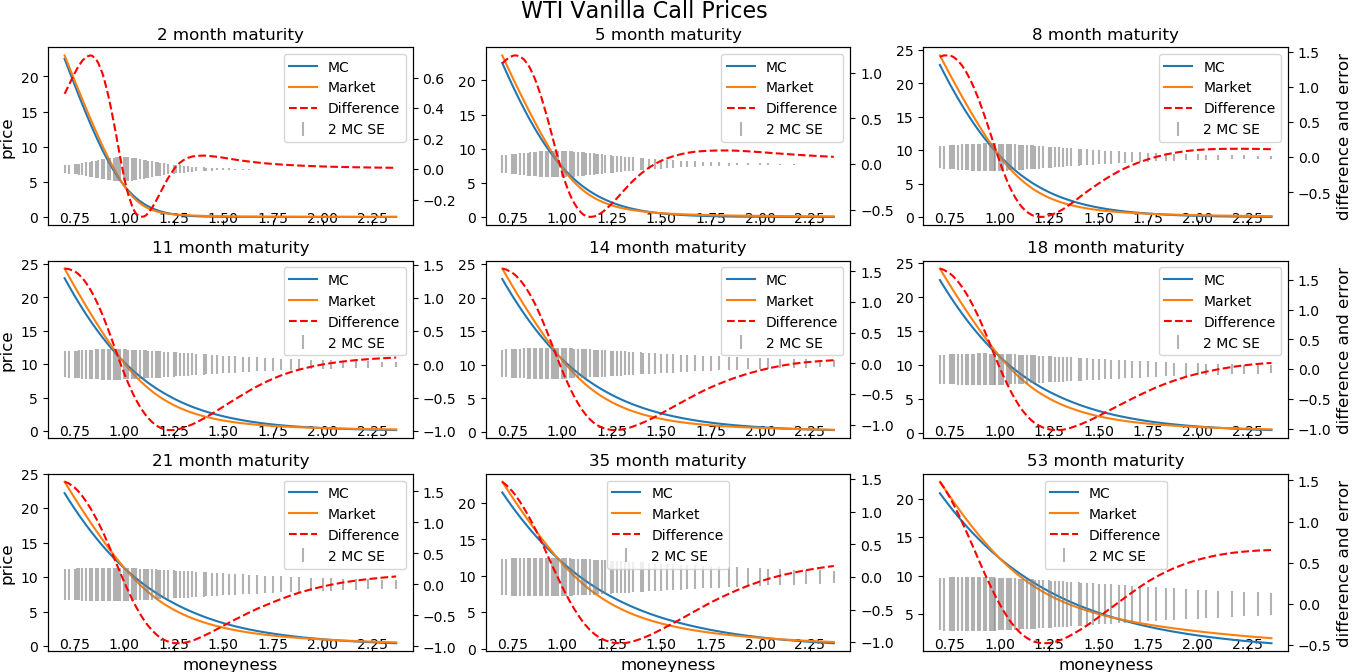}
    \newline\vspace{5pt}
    \includegraphics[width=\textwidth]{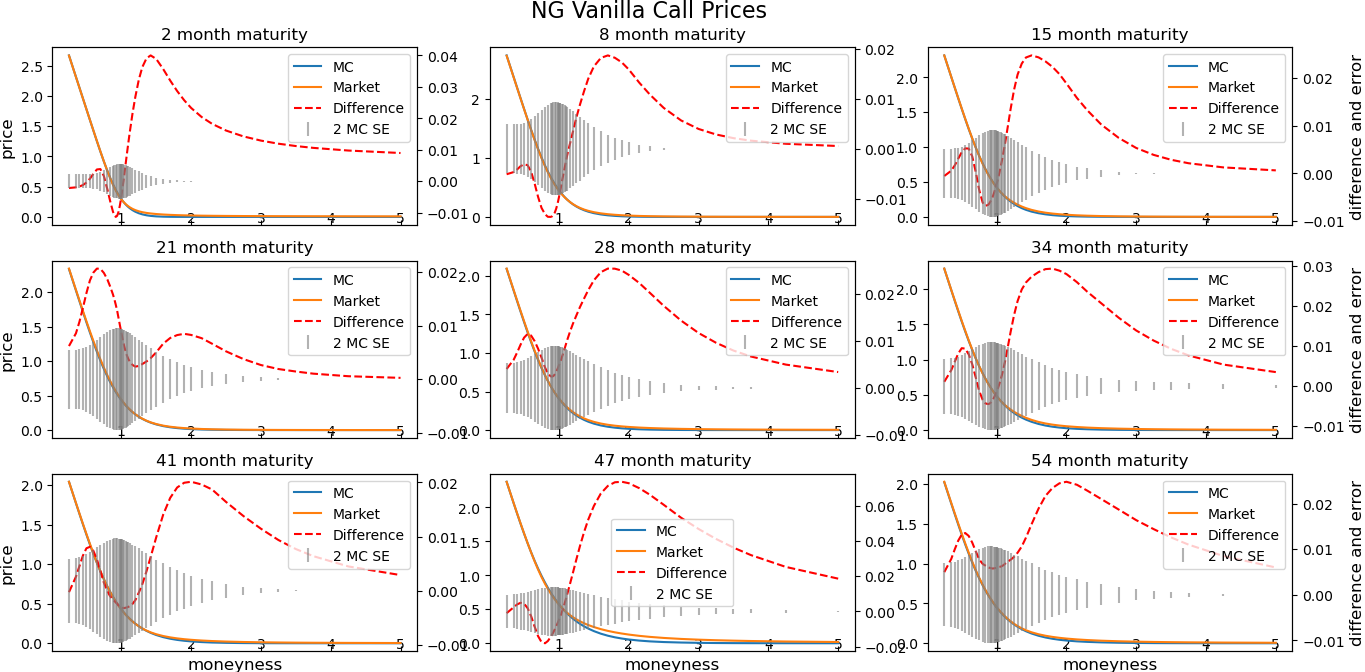}
    \caption{Monte Carlo prices compared to market prices for WTI and NG for
    various maturities and moneynesses}
    \label{fig:call_prices_Andersen}
\end{figure}

Next we simulate the same model to price vanilla call options at various
moneynesses and maturities. Here we compare Monte Carlo prices to market prices
as given by market implied volatility data. Figure
\ref{fig:call_prices_Andersen} shows that while the model performs well at
at-the-money region (around moneyness $K / F_j(0) = 1$), the differences to
market prices are significantly greater than Monte Carlo standard errors in most
other regions.
% Script used:
% scripts/WTI_simulate_Andersen.ipynb scripts/NG_simulate_Andersen.ipynb

\FloatBarrier

\section{Modeling the Smile}\label{sec:Andersen_LV}
While the above model reproduces the at-the-money implied volatilities,
it performs poorly on the wings of the WTI and NG implied volatility surfaces.
The original work of Andersen proposes an extension with a regime switching
jump model with parameters approximately fit to capture some skew information 
from the market and/or desk volatility smile.
Here we take a non-parametric local volatility approach.

\subsection{The Leverage Functions}

We generalize the model \eqref{eqn:andersen_sde} as
\begin{equation}
dF_j(t) = F_j(t) L_j\left(F_j(t), t\right) \left[\sigma_1(t, T_j) dW_1(t) +
\sigma_2(t, T_j) dW_2(t) \right].\label{eqn:andersen_localvol_sde}
\end{equation}
We formulate the model with unique leverage functions $L_j(K, t)$ for each
futures delivery. 
For each futures delivery, there is only at most one standard liquidly 
quoted and traded option maturity $T_j$ with maturity during the time when
this futures contract will be the next one to be delivered.\footnote{For 
some delivery months, liquid option trading for the 
standard maturity might not have started yet and hence those contracts 
might not have liquid volatility data yet. In those circumstances, we use 
an accumulator derived from the implied volatility of 
the next liquidly traded longer contract.}
Thus, only a single time slice $T_j$ of implied volatility
data $\Sigma_j(K, T_j)$ will be available for $F_j(t)$.
The leverage functions $L_j(K, t)$ will be calibrated to this data.

We assume that the num\'eraire associated with the risk-neutral measure
$\mathbb{Q}$, that is the money market account $B(t)$, accrues at short rate
$r(t)$ by $dB(t) = r(t) B(t) dt$. The short rate is modeled by a single factor
process of generic form
\begin{equation}
dr(t) = \alpha(\omega, t) dt + \sigma(\omega, t)
dW^r(t),\label{eqn:generic_r_sde}
\end{equation}
where $\alpha$ and $\sigma$ are bounded functions of a general set of
stochastic factors $\omega \in \Omega$, and $W^r$ is a Brownian motion under
risk-neutral measure $\mathbb{Q}$. The discount factor is given
by $D(t)\equiv 1/B(t)=\exp{\left[-\int_0^tr(u) du\right]}$.

We denote by $P(t, T) \equiv
\mathbf{E}^{\mathbb{Q}} \left[\frac{D(T)}{D(t)} \middle\vert \mathcal{F}_t
\right]$ the time $t$ value of the zero coupon bond maturing at time $T$,
through which we define the instantaneous forward rate
\begin{equation}
f(t, T) \equiv - \frac{\partial \log P(t, T) }{\partial T}
= - \frac{1}{P(t, T) }\frac{\partial P(t, T) }{\partial T}.
\end{equation}

Let $C_j=C_j(K, t)$ denote the time-zero price of a vanilla call option with
strike $K$ and expiry $t\leq T_j$ written on $F_j$. In the vanilla call option
price formulation, the leverage functions for the futures with delivery $T_j$
are given by \cite{Ogetbil2020}
\begin{equation}
L_j(K, t)^2 = \frac{\frac{\partial C_j}{\partial t} +
\mathbf{E}^{\mathbb{Q}}\left[D(t) (F_j(t) - K) r(t) \mathds{1}_{F_j(t) > K}\right]
}{\frac{1}{2}K^2\frac{\partial^2 C_j}{\partial K^2} \left[\sigma_1(t, T_j)^2 +
\sigma_2(t, T_j)^2\right]},\label{eqn:andersen_localvol_full}
\end{equation}
where the expectation is taken under risk-neutral measure.

\subsection{Total Implied Variance Accumulation}
Evaluating the call option price derivative above in time direction is not
trivial as the market quotes data comes in a single time slice. The total
implied variance, on the other hand, grows in time and has value zero
at time zero, and hence is more straightforward to differentiate in time
direction.

It is convenient to parametrize the total implied variance $w_j$ for the vanilla
options expiring at time $t$, written on spot futures delivered at time $T_j$ in
terms of \emph{log-moneyness}
\begin{equation}
y_j(K, t) \equiv \log\frac{K}{F_j(t)}.
\end{equation}
We assume that the input implied volatility surface is transformed into
log-moneyness coordinates as $\Sigma_j(y, t)$. Here we use the same
log-moneynesses at all maturities and we drop the subscript $j$ from $y_j$.
The total implied variance is given by
\begin{equation*}
w_j(y, t) \equiv \Sigma_j(y, t)^2 t,
\end{equation*}
where $\Sigma_j(y, t)$ is the implied volatility of a vanilla option expiring at
time $t\leq T_j$ struck at $K = F_j e^{y}$, written on $F_j$. By means of the
available implied volatility data $\Sigma_j(y, T_j)$ we compute
$\tilde{w}_j(y) \equiv w_j(y, T_j)$ for all $y$ given in the data set. We also
know $w_j(y, 0) = 0$.
The total implied variance is monotonically increasing between the two slices,
but otherwise we have 
%some freedom in between while creating a total implied
%variance accumulator.
control over how quickly the total implied variance will accumulate toward its
given final value. 

A straightforward approach would be to model accumulation at a constant rate
in time direction, creating a linear accumulator as 
\begin{equation}
w_j(y, t) = \Sigma_j(y, T_j)^2 t = \tilde{w}_j(y)
\frac{t}{T_j},\ 0 \leq t \leq T_j\ \ \ \ \text{(linear accumulator)}
\label{eq:linear_acc}
\end{equation}
With a quadratic function, the total implied variance accumulates faster towards
delivery,
\begin{equation}
w_j(y, t) = \tilde{w}_j(y)
\left(\frac{t}{T_j}\right)^2,\ 0 \leq t \leq T_j\ \ \ \ \text{(quadratic
accumulator)}
\label{eq:quadratic_acc}
\end{equation}

As an alternative, we can use the implied volatility from options of futures
closer to their maturity which would correspond to the assumption that options
with the same time-to-maturity should have the same or similar implied
volatility.
Thus, when the $T_j$ futures option is $T_k$ ($k<j$) away from its maturity (at
time $T_j-T_k$), its remaining total implied variance should be the total
implied variance from  the $T_k$ futures option.
Thus, we propose that from time $T_j - T_k, k \leq j$ to time $T_j$ a total
implied variance of $\tilde{w}_k(y) \equiv \Sigma_k(y, T_k)^2 T_k$ accumulates.
With this assumption, the total implied variance is $\tilde{w}_j(y) -
\tilde{w}_k(y)$ at time $T_j - T_k$.
Setting $T_0 \equiv \tilde{w}_0(y) \equiv 0$, we create a piecewise linear total
implied variance interpolator at times
% $0, \ldots, T_j - T_2, T_j - T_1 , T_j$
$\{T_j - T_k; k = j, \ldots, 0\}$ and values
% $0, \ldots, \tilde{w}_j(y) - \tilde{w}_2(y), \tilde{w}_j(y) - \tilde{w}_1(y),
% \tilde{w}_j(y)$
$\{\tilde{w}_j(y) - \tilde{w}_k(y); k = j, \ldots, 0\}$ as
\begin{equation}
w_j(y,t) =
\tilde{w}_j(y) - \tilde{w}_k(y) + \frac{\tilde{w}_k(y) - \tilde{w}_{k-1}(y)}{T_k
- T_{k-1}} (t-T_j+T_k),\ 0 \leq t \leq T_j\ \ \ \text{(TTM IV accumulator)}
\label{eq:ttmiv_acc}
\end{equation}
where $k = \argmin_{i \in \{1, \ldots, j\}}\{T_j - T_i \leq t\}$.
The procedure is repeated for every target delivery $T_j$ so that each delivery
has its own total implied variance accumulator.
% Script used:
% scripts/tiv_accumulation.py
\begin{figure}[ht!]
    \centering
    \includegraphics[width=1\textwidth]{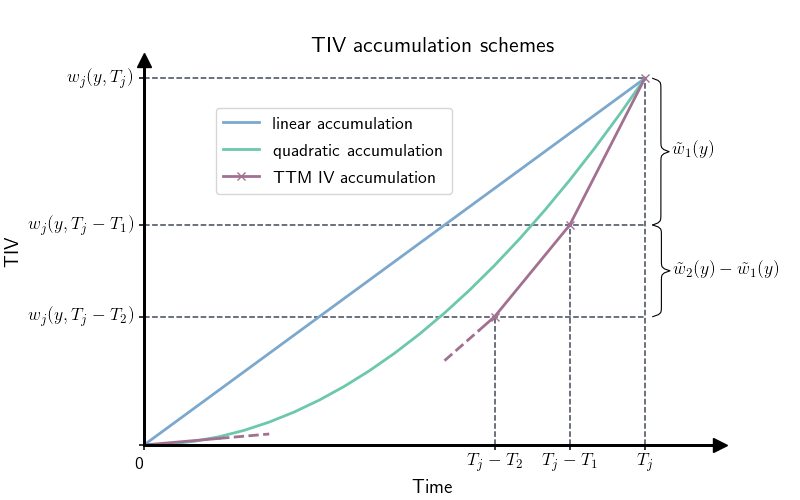}
    \caption{Different approaches to accumulate $w_j(y,t)$ at a given $y$
    along time dimension between times 0 and $T_j$, and values 0 and
    $\tilde{w}_j(y)=w_j(y, T_j)$. In the market mimicking accumulation scheme,
    one uses volatility data $\tilde{w}_k(y), k < j$ from previous slices.}
    \label{fig:tiv_accumulation}
\end{figure}
%\FloatBarrier
When implementing this mimicking
approach one needs to make sure that the total implied variance remains positive
and monotonically increasing. If the short term futures volatilities are
sufficiently high, it is possible that the total implied variance only starts to
accumulate close to the maturity for some futures. Figure
\ref{fig:tiv_accumulation} sketches the three schemes. 

Instead of taking implied volatility from futures option with shorter time-to-maturity 
as interpolation targets, one could introduce weighting factors $W_{k,j}$ or functions, either associated with 
option maturity times $T_k$ or fixed times to maturity $T_{j,k}=T_j-T_k$ with $w_j(y,T_k) = W_{k,j} 
w_j(y,T_j)$ or  $w_j(y,T_k) = W_{k,j} w_j(y,T_{j,k})$, controlling the accumulation of the 
total implied variance without changing or controling the shape and skew. 
While there is some disagreement about how to measure the Samuelson maturity effect, 
there seems to be agreement that at least for certain commodity futures (agricultural 
and energy-related), there is in general larger volatility (in the sense of realized quadratic 
variance or volatility) or range of those futures prices (agricultural and energy) 
the closer the delivery of the futures (and the maturity of the option) is\footnote{See, 
for instance, \cite{jaeck2014samuelson} or \cite{daalreexamining}.
}. This would correspond to small and slowly increasing 
weighting factors when the option is far from its maturity and to larger and faster 
increasing weighting factors closer to its maturity. Similarly,
one could replace $\frac{t}{T_j}$ in (\ref{eq:linear_acc}) by a function $f(\frac{t}{T_j})$
with $f(0)=0$ and $f(1)=1$ that expresses the maturity effect, like
$f(x)=\frac{e^x - 1}{e - 1}$.

Finally, one could take a mixture of the above strategies. In this, 
a variety of maturity effects can be flexibly implemented while maintaining 
calibration to the given option skew data.

\begin{comment}
 This allows us to create a total implied variance interpolator, piecewise cubic
 in log-moneyness direction, for $0 < t \leq T_j$ as
\begin{equation}
w_j(y_j, t) = \Sigma_j^2(K, T_j) t \frac{u(t)}{u(T_j)}, \label{eqn:tiv_weights}
\end{equation}
where we introduce the ``weight function'' $u(t)$ for total implied variance at
time $t$, subject to the constraint that the total implied variance $w_j(y_j,
t)$ is monotonically increasing in time. The simple case with $u(t) \equiv 1$
corresponds to simple linear interpolation of the total implied variance along
the time direction. As an alternative, to capture the volatility behavior at
short maturities we consider the at-the-money total implied variance curve. This
curve has values $w_0 \equiv 0, w_1 \equiv \Sigma_1^2(F_1(0), T_1) T_1, w_2
\equiv \Sigma_2^2(F_2(0), T_2) T_2, \ldots$ at times $0, T_1, T_2, \ldots$
Through the piecewise linear interpolator of this curve
\begin{equation}
w(t) = \frac{w_{i+1} (t - T_i) + w_i (T_{i+1} - t)}{T_{i+1} - T_i},
\begin{cases}
    i=T_i=0,& \text{if } t < T_1 \\
    i=\max\{j | T_j \leq t \}, & \text{otherwise}
\end{cases}
\end{equation}
we can define the weight function $u(t) \equiv \frac{w(t)}{t}$ for $t > 0$.
This function allows the total implied variance $w_j(y_j, t)$ mimick the
at-the-money volatility behavior of the short term futures at earlier times,
that is $w_j(0, T_k) = w_k(0, T_k)$ for all $k \leq j$.
\end{comment}

In log-moneyness parametrization, the Black-Scholes European call option price
function $C_j$ can be written as
\begin{equation}
C_j(P(0, t) F_j(0), y, w_j(y, t)) = P(0, t) F_j(0) \left(N(d_1) - e^{y}
N(d_2)\right)\label{eqn:C_BS}
\end{equation}
with
\begin{equation*}
\begin{split}
d_1 \equiv& -y w_j^{-\frac{1}{2}} + \frac{1}{2} w_j^{\frac{1}{2}},\\
d_2 \equiv& d_1 - w_j^{\frac{1}{2}},
\end{split}
\end{equation*}
and $N(\cdot)$ is the cumulative Gaussian probability distribution function.
The derivatives of the call option price are computed as
\begin{equation}
\begin{split}
\frac{1}{2}K^2 \frac{\partial^2 C_j}{\partial K^2} =&
\frac{\partial C_j}{\partial w_j} \left[
1 - \frac{y}{w_j} \frac{\partial w_j}{\partial y}
+ \frac{1}{2} \frac{\partial^2 w_j}{\partial y^2}
+ \frac{1}{4} \left(\frac{\partial w_j}{\partial y}\right)^2
\left(-\frac{1}{4}- \frac{1}{w_j} + \frac{y^2}{w_j^2}\right)
\right],\\
\frac{\partial C_j}{\partial t} =& \frac{\partial C_j}{\partial
w_j}\frac{\partial w_j}{\partial t} - f(0, t) C_j,
\end{split}\label{eqn:callderivs}
\end{equation}
where
\begin{equation}
\frac{\partial C_j}{\partial w_j} = \frac{1}{2}P(0, t)F_j(0) e^{y} N'(d_2)
w_j^{-\frac{1}{2}}.\label{eqn:dCdw}
\end{equation}
Plugging the quantities \eqref{eqn:callderivs} into the identity for the
leverage functions \eqref{eqn:andersen_localvol_full}, we find
\begin{equation}
L_j(y, t)^2 = \frac{\frac{\partial C_j}{\partial w_j}\frac{\partial
w_j}{\partial t} - f(0, t) C_j + 
\mathbf{E}^{\mathbb{Q}}\left[D(t) F_j(t) (1 - e^y) r(t) \mathds{1}_{y<0}
\right]} {\frac{\partial C_j}{\partial w_j}\left[1 - \frac{y}{w_j} \frac{\partial
w_j}{\partial y} + \frac{1}{2} \frac{\partial^2 w_j}{\partial y^2}
+ \frac{1}{4} \left(\frac{\partial w_j}{\partial y}\right)^2
\left(-\frac{1}{4}- \frac{1}{w_j} + \frac{y^2}{w_j^2}\right)\right]
\left[\sigma_1(t, T_j)^2 + \sigma_2(t, T_j)^2\right]},\label{eqn:lev_tiv_sto_r}
\end{equation}
with $C_j$ and $\frac{\partial C_j}{\partial w_j}$ as given in
\eqref{eqn:C_BS} and \eqref{eqn:dCdw}.
The above expression can be used to compute leverage functions $L_j(y, t)$ for
each futures underlier $F_j(t)$ iteratively in a bootstrapping fashion.

\paragraph{Deterministic rate limit}

In this case we have $\alpha(\omega, t)=\partial f(0, t) / \partial
t, \sigma(\omega, t)=0, r(t) = f(0, t)$, and \eqref{eqn:andersen_localvol_full}
simplifies to
\begin{equation}
L_j(K, t)^2 = \frac{\frac{\partial C_j}{\partial t} + f(0, t) C_j
}{\frac{1}{2}K^2\frac{\partial^2 C_j}{\partial K^2} \left[\sigma_1(t, T_j)^2 +
\sigma_2(t, T_j)^2\right]}.\label{eqn:andersen_C_det_r}
\end{equation}
In the total implied variance formulation this can be written as
\begin{equation}
L_j(y, t)^2 = \frac{\frac{\partial w_j}{\partial t}}
{\left[1 - \frac{y}{w_j} \frac{\partial w_j}{\partial y}
+ \frac{1}{2} \frac{\partial^2 w_j}{\partial y^2}
+ \frac{1}{4} \left(\frac{\partial w_j}{\partial y}\right)^2
\left(-\frac{1}{4}- \frac{1}{w_j} + \frac{y^2}{w_j^2}\right)\right]
\left[\sigma_1(t, T_j)^2 +
\sigma_2(t, T_j)^2\right]}.\label{eqn:lev_tiv_det_r}
\end{equation}

\subsection{Calibrating the Leverage Functions}
We adapt the calibration approach proposed in \cite{OGH2022} to compute the
leverage functions $L_j(\cdot, t)$ for every futures delivery $F_j(t)$
simultaneously time slice by time slice. In the deterministic rate case,
the calibration is straightforward since all quantities in
\eqref{eqn:lev_tiv_det_r} are known. For stochastic rate, we perform a Monte
Carlo simulation to estimate the expectation appearing in
\eqref{eqn:lev_tiv_sto_r}.

\paragraph{Inputs for calibration}
Our calibration routine expects the following quantities as input for leverage
function calibration:
\begin{itemize}
  \item Andersen model \eqref{eqn:andersen_sde} parameters calibrated to market
  data
\item Futures prices $F_j(0)$ as of the valuation time $t=0$
\item Market implied volatility $\Sigma_j(y, T_j)$ for each maturity
\item Market yield curve $P(0, T)$
\item For a stochastic rate model, 
%If modeling the rate stochastically, 
a short rate model with parameters calibrated to market data.
%\item If modeling the rate stochastically, 
\item For a stochastic rate model, 
coefficients of correlation between
the short rate and the two Brownian motions of the Andersen model
\end{itemize}

\paragraph{Steps for calibration}

We calibrate the leverage functions time slice by
time slice, in a bootstrapping fashion. Let $t_i; i=1, \ldots, n$ be the
increasing sequence of (positive) times where we will perform the calibration.
\begin{enumerate}
  \item Using the market implied volatilities $\Sigma_j(y, T_j)$, generate a
  total implied variance
  surface $w_j(y, t)$ interpolator. The interpolator must be able to compute
  the partial derivatives appearing in the local volatility expressions.
  \item For the first time slice $t_1$, evaluate the
  deterministic equation \eqref{eqn:lev_tiv_det_r}
  to compute the leverage function values $L_j(y, t_1)$ for a predetermined
  range of strikes for every $j$.
  This step requires no Monte Carlo simulation. As a result, obtain leverage
  function values to be used until time $t_2$ in the subsequent calibration
  steps.
  \item For each of the subsequent time slices $t_i, i > 1$, (a)
  deterministic rate case: evaluate \eqref{eqn:lev_tiv_det_r}
  directly for a predetermined
  range of strikes for every $j$, (b) stochastic rate case: Simulate the SDE
  system \eqref{eqn:andersen_localvol_sde}, \eqref{eqn:generic_r_sde} up to time
  $t_i$. Compute the Monte Carlo estimate for the expectation appearing in
  \eqref{eqn:lev_tiv_sto_r} for a predetermined range of strikes for every $j$.
  Use these equations to obtain the leverage function values  $L_j(y, t_i)$.
  These values will be used during subsequent simulation steps from time $t_i$
  to time $t_{i + 1}$. This step is first performed with $i=2$ and is then
  repeated for the remaining time slices.
\end{enumerate}
The leverage functions strike grid is chosen uniformly in a region covered by
the implied volatility data.

\subsection{Implementation Tests}

\begin{figure}[ht!]
    \centering
    \includegraphics[width=.32\textwidth]{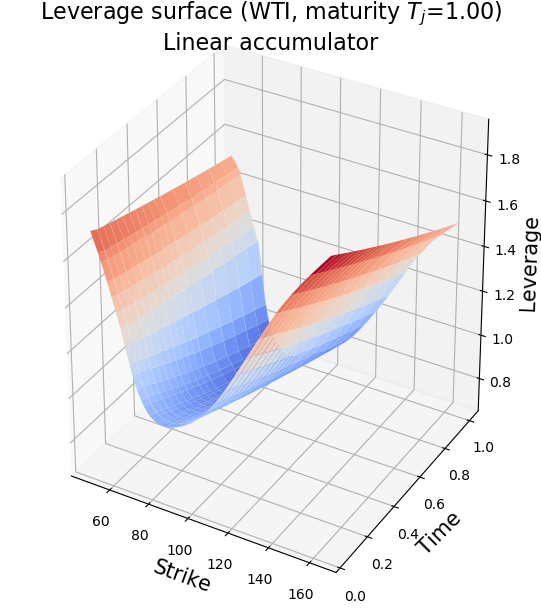}
    \includegraphics[width=.32\textwidth]{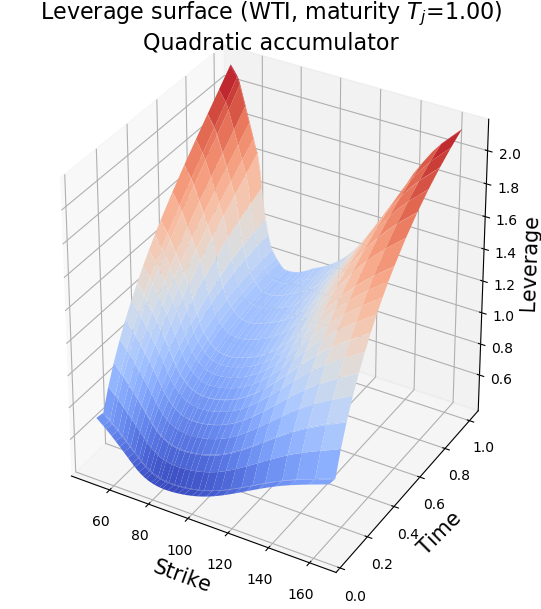}
    \includegraphics[width=.32\textwidth]{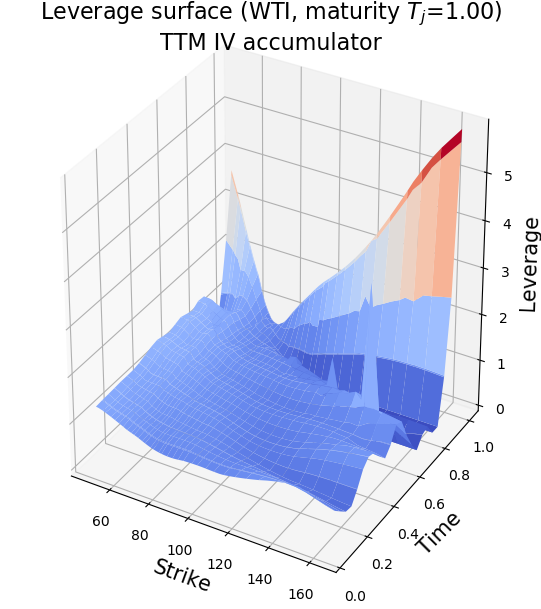}
    \caption{Leverage functions, computed with linear accumulation
    between times 0 and $T_j$ (left), computed with quadratic accumulation
    (center), computed with nonlinear accumulation
    based on total implied variances from option prices with shorter 
    time-to-maturity (right).}
    \label{fig:leverage_AndersenLV_WTI}
\end{figure}

We apply the above recipe to market data for WTI and NG as of 2021-12-31 and
compute the leverage functions $L_j(y, t)$. In Figure
\ref{fig:leverage_AndersenLV_WTI}, we observe the leverage functions under
linear \eqref{eq:linear_acc}, quadratic \eqref{eq:quadratic_acc}, and TTM IV
\eqref{eq:ttmiv_acc} accumulation schemes described in the previous subsection,
for the WTI futures underlier with 12-month delivery. As expected, the leverage
function is smooth in the linear case whereas most of the magnitude and
variation of the leverage function occurs closer to maturity in the nonlinear
cases.

% Script used:
% scripts/WTI_calibrate_localvol.ipynb
% scripts/WTI_calibrate_localvol_newtiv2.ipynb
% scripts/WTI_calibrate_localvol_quadratic.ipynb

\begin{figure}[ht!]
    \centering
    \includegraphics[width=1\textwidth]{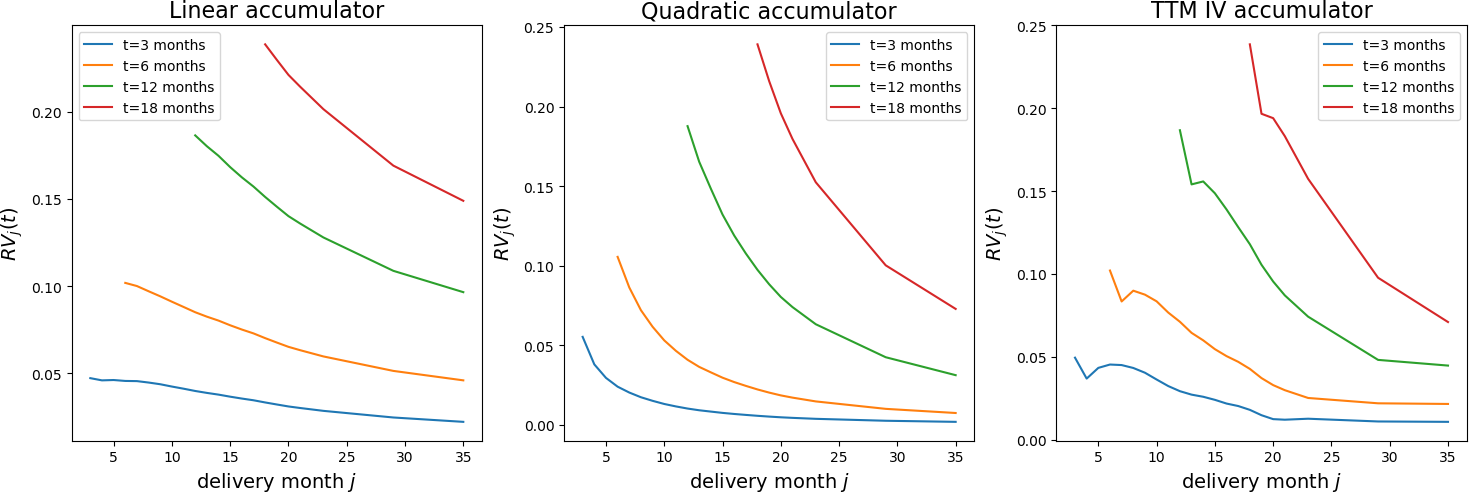}
    \caption{Realized variance $RV_j(t)$ for several $t$ as a function of
    delivery month $j$ estimated by Monte Carlo simulation with 10000 paths,
    using linear, quadratic, and TTM IV accumulators.}
    \label{fig:realized_var2_AndersenLV_WTI}
\end{figure}

\begin{figure}[ht!]
    \centering
    \includegraphics[width=1\textwidth]{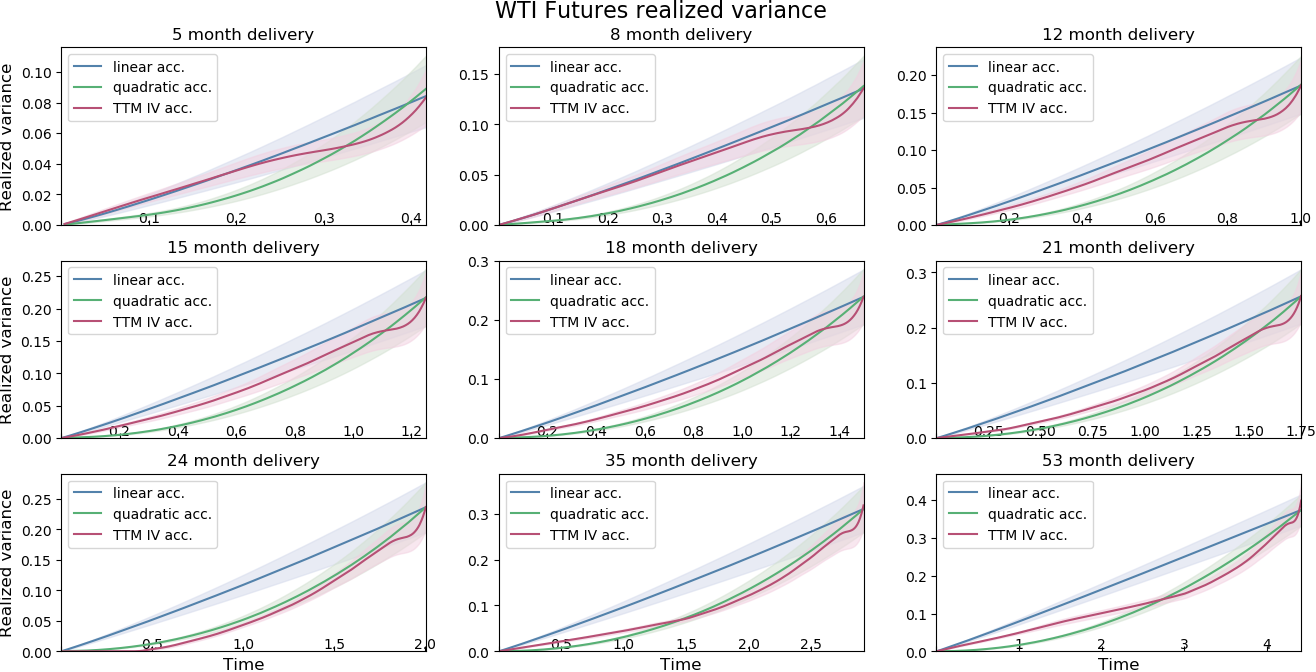}
    \caption{Realized variance $RV_j(t)$ for several deliveries $T_j$ with
    linear, quadratic, and TTM IV accumulators. Solid lines represent the Monte
    Carlo mean for 10000 paths, and the shaded areas span one standard error
    around the mean.}
    \label{fig:realized_var_AndersenLV_WTI}
\end{figure}

To investigate the variance accumulation of the underlying futures, we simulate
the calibrated model on a time partition $\{t^{(j)}_s; s=0, \ldots, m_s\}$ with
$t^{(j)}_0=0; t^{(j)}_{s-1} < t^{(j)}_{s};$ and $t^{(j)}_{m_s} = T_j$ for
several deliveries $T_j$ over 10000 paths. We compute the Monte Carlo estimate
and error of the realized variance $RV_j(t)\equiv \sum_s^{t_s=t}\left[\log
\frac{F_j(t_s)}{F_j(t_{s-1} )}\right]^2$ for the simulated model.
In Figure \ref{fig:realized_var2_AndersenLV_WTI} we plot the realized variance
as a function of delivery month for several $t$s. One observes that the realized
variance increases with $t$. Moreover the realized variance is higher when $t$
is near the delivery $T_j$, also steeper for nonlinear accumulators near the
delivery; and it decreases in general as the delivery is further from $t$.
% Script used:
% scripts/WTI_simulate_localvol_realized_var.ipynb
Figure \ref{fig:realized_var_AndersenLV_WTI} shows the realized variance for the
three total implied variance accumulation approaches we considered. As we
expected, the methodologies result in the same realized variance at the
delivery up to Monte Carlo estimation errors. We also see that the total
implied variance accumulates faster close to the delivery in the nonlinear
(quadratic, TTM IV) cases, consistent with the market implied volatility data
(see Figure \ref{fig:atm_vols}) and the differences in the intermediate months
are more prominent on futures with longer deliveries.
% Script used:
% scripts/WTI_simulate_localvol_realized_var.ipynb
\FloatBarrier

For the stochastic rate case, we assume that the domestic short rate follows a
G1++ process\footnote{G1++ can be seen as a reparametrization of Hull-White or 
LGM models. Any other calibrated interest model could also be used as long as 
short rate and its time integral can be simulated.}
\begin{equation}
\begin{split}
r(t) &= x(t) + \phi_t, \\
dx(t) &= -a x(t) dt + \sigma dW^{r}(t),
\label{eqn:dr_G1PP}
\end{split}
\end{equation}
where $W^{r}$ is a Brownian motion under risk-neutral measure $\mathbb{Q}$,
$\phi_t$ is the shift function that ensures
$P(0, t) = \mathbf{E}^{\mathbb{Q}}\left[D(t)\right]$, $a$ is the mean reversion
coefficients, and $\sigma$ is the volatility coefficient. In our tests, the
parameters are chosen as $a=0.02$, $\sigma = 0.01$, and the coefficients of
correlation between the Brownian motions, defined as in $d\left<W^{1}, W^{r}
\right>_t = \rho_{1r} dt$, and $d\left<W^{2}, W^{r} \right>_t = \rho_{2r} dt$,
are set as $ \rho_{1r} = \rho_{2r} = -0.2$. The expectation appearing in
\eqref{eqn:lev_tiv_sto_r} is estimated by simulating 1000 Monte Carlo paths and
their antithetic conjugates. We calibrate the WTI leverage functions with the
TTM IV accumulator \eqref{eq:ttmiv_acc}, whereas for calibrating the NG leverage
functions we choose the quadratic accumulator \eqref{eq:quadratic_acc}.

% Script used:
% scripts/WTI_simulate_localvol.ipynb
% scripts/NG_simulate_localvol_quadratic.ipynb
\begin{figure}[ht!]
    \centering
    \includegraphics[width=\textwidth]{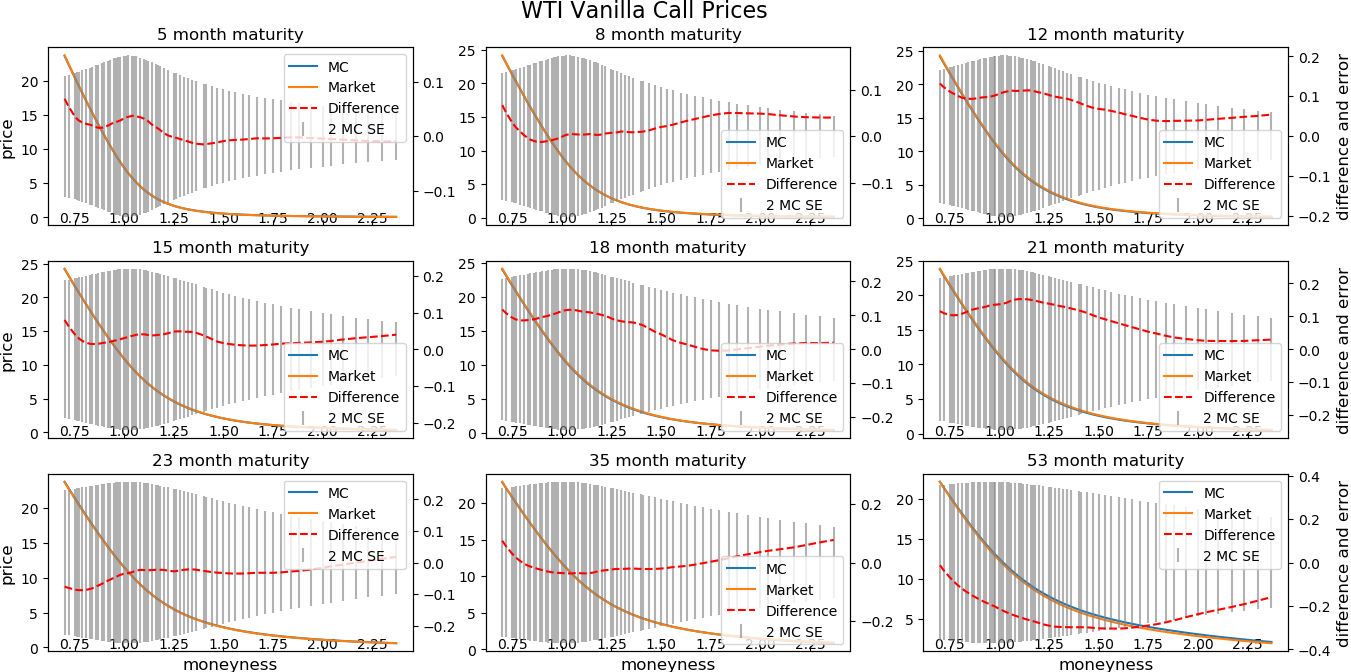}
    \newline\vspace{5pt}
    \includegraphics[width=\textwidth]{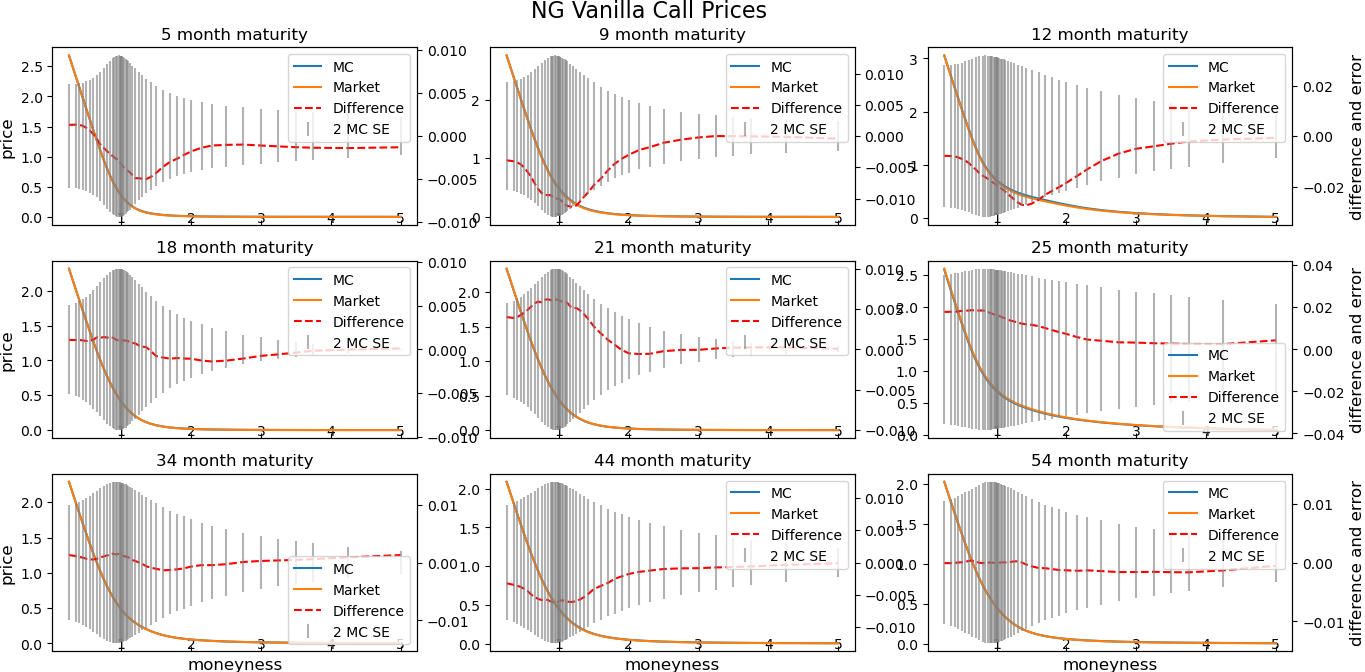}
    \caption{Market and Monte Carlo prices for vanilla call options on WTI and
    NG futures at various maturities. Rate is deterministic. We simulate 10000
    paths and their antithetic conjugates to compute the Monte Carlo estimates
    and standard errors for prices.}
    \label{fig:call_prices_AndersenLV}
\end{figure}

The calibrated leverage functions are validated by simulating the model,
with 10000 paths and their antithetic conjugates, to price vanilla call options
at various moneynesses and maturities.
For the deterministic rate case, Figure \ref{fig:call_prices_AndersenLV} shows that the
market prices are within two Monte Carlo standard errors of the simulated prices
for most strikes in the test range.
% Script used:
% scripts/WTI_simulate_localvol_g1pp_newtiv2.ipynb
% scripts/NG_simulate_localvol_g1pp_quadratic.ipynb
\begin{figure}[ht!]
    \centering
    \includegraphics[width=\textwidth]{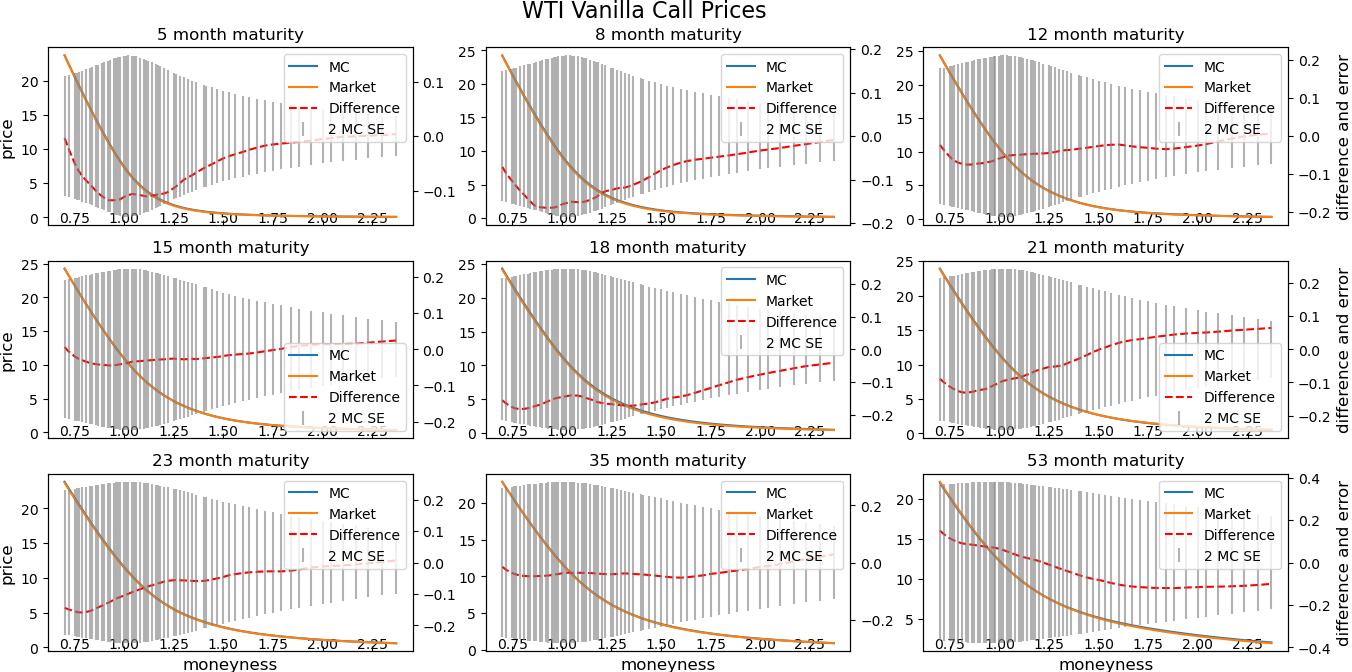}
    \newline\vspace{5pt}
    \includegraphics[width=\textwidth]{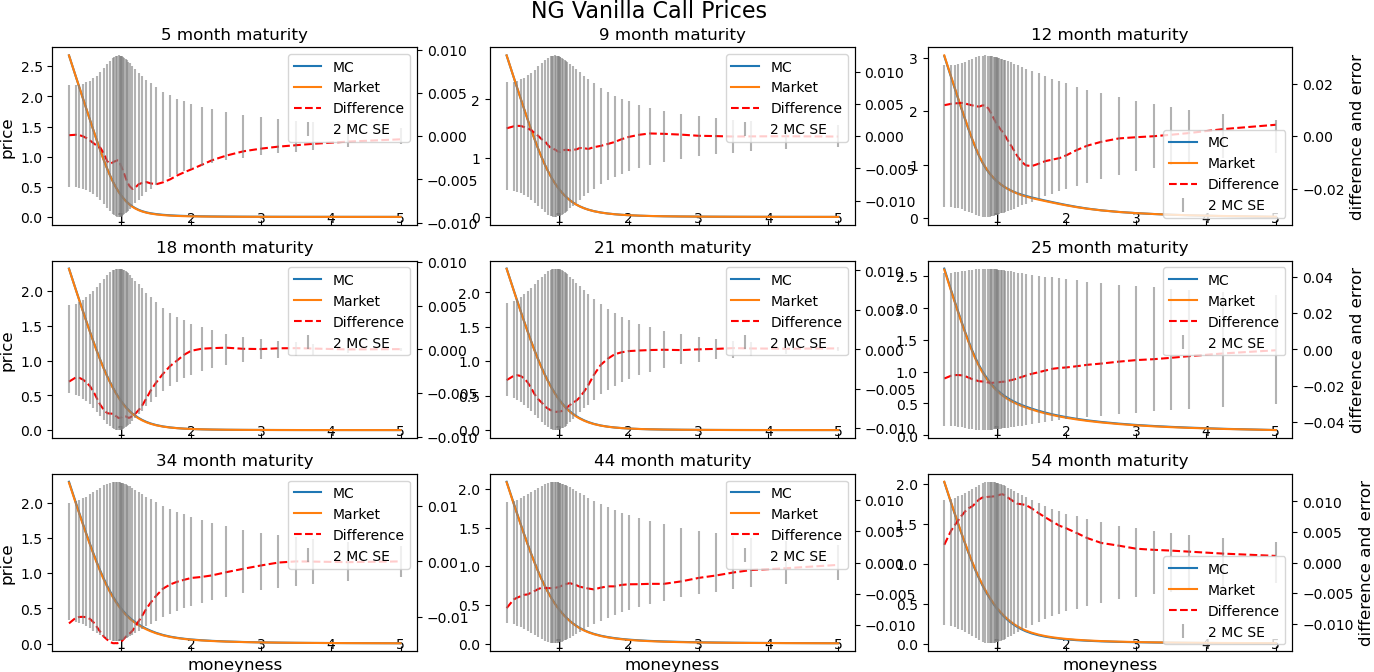}
    \caption{Market and Monte Carlo prices for vanilla call options on WTI and
    NG futures at various maturities. The stochastic short rate follows a G1++
    process.     We simulate 10000 paths and their antithetic conjugates to
    compute the Monte Carlo estimates and standard errors for prices.}
    \label{fig:call_prices_AndersenLV_g1pp}
\end{figure}
Similarly for the stochastic rate case, as shown in Figure
\ref{fig:call_prices_AndersenLV_g1pp}, the market prices are within two Monte
Carlo standard errors of the simulated prices.

These results show that the calibrated leverage functions recover the market
volatility smile well, and that the stochastic rate does not introduce
significant inaccuracy to the model.

\section{Discussion}\label{sec:discussion}
The demonstration of price recovery shows that the calibrated model
captures WTI and NG market volatility smiles in both deterministic and
stochastic rate cases. While remaining in the risk-neutral framework, the total
implied variance for the entire maturity determined during the model calibration from market
data can be accumulated over time in different ways. For instance, one can use 
the total implied variance from other futures options with shorter time-to-maturity 
to define the accumulation over the longer option maturity, reflecting information 
from all implied volatility slices. This can result in models where the total implied
volatility accumulates faster than linearly close to the corresponding futures option
maturity, which is commonly referred to as the Samuelson or maturity effect. We 
also mentioned other ways to directly control the accumulation of total 
implied variance.
%way to mimick market structure, where options are
%observed to have increased volatility when approaching their maturity, which is
%commonly referred to as the Samuelson effect. 
Further studies showed that the
market prices are recovered comparably well to the linear interpolation case
when the total implied variance is interpolated with the suggested methodology
which gives rise to higher realized volatilities towards maturity.

One can further generalize the smile dynamics by adding stochasticity to the
diffusion process \eqref{eqn:andersen_localvol_sde},
\begin{equation}
dF_j(t) = F_j(t) \sqrt{U_j(t)} L_j\left(F_j(t), t\right) \left[\sigma_1(t, T_j)
dW_1(t) + \sigma_2(t, T_j) dW_2(t)
\right],\label{eqn:andersen_stochastic_localvol_sde}
\end{equation}
where $U_j(t)$ are adapted processes that model the stochastic variance. A
typical choice would be the CIR process \cite{CIR1985} with the $U_j(t)$ process
parameters calibrated to near at-the-money vanilla options with maturity $T_j$.
For this extended model, the leverage functions are given by \cite{Ogetbil2020}
\begin{equation}
L_j(K, t)^2 = \frac{P(t, T_j)\left(\frac{\partial C_j}{\partial t} +
\mathbf{E}^{\mathbb{Q}}\left[D(t) (F_j(t) - K) r(t) \mathds{1}_{F_j(t) >
K}\right]\right) }{\frac{1}{2}K^2\frac{\partial^2 C_j}{\partial K^2}
\left[\sigma_1(t, T_j)^2 + \sigma_2(t, T_j)^2\right] \mathbf{E}^{\mathbb{Q}}\left[D(t)
U_j(t) | F_j(t) = K \right] },\label{eqn:andersen_stochastic_localvol_full}
\end{equation}
where the expectations are taken under risk-neutral measure $\mathbb{Q}$. The
calibration of the leverage functions proceeds as before. Only this time, one
additionally needs to compute the value of the conditional expectation appearing
in the above expression. Since this is a multi-dimensional problem, one can
employ Monte Carlo methods, such as the binning or the regression based
techniques proposed in \cite{OGH2022} to estimate the value of this conditional
expectation.

\paragraph{Acknowledgments}
The authors thank Agus Sudjianto for supporting this research, and Vijayan
Nair for his comments and suggestions regarding this research. Any opinions,
findings and conclusions or recommendations expressed in this material are those
of the authors and do not necessarily reflect the views of Wells Fargo Bank,
N.A., its parent company, affiliates and subsidiaries.

%\newpage{}

\bibliographystyle{unsrt}
\bibliography{commodity_lv}

\begin{thebibliography}{10}

\bibitem{Gabillon1991}
J.~Gabillon.
\newblock {\em The term structure of oil futures prices}.
\newblock OIES: M17. Oxford Institute for Energy Studies, 1991.

\bibitem{Andersen2010}
Leif Andersen.
\newblock {Markov models for commodity futures: theory and practice}.
\newblock {\em Quantitative Finance}, 10(8):831--854, 2010.

\bibitem{NPS2020}
Emanuele Nastasi, Andrea Pallavicini, and Giulio Sartorelli.
\newblock {Smile Modeling In Commodity Markets}.
\newblock {\em International Journal of Theoretical and Applied Finance
  (IJTAF)}, 23(03):1--28, May 2020.

\bibitem{MNPV2022}
Alberto Manzano, Emanuele Nastasi, Andrea Pallavicini, and Carlos V\'azquez.
\newblock Pricing commodity index options, 2022.
\newblock arXiv:2208.01289.

\bibitem{schneider2018seasonal}
Lorenz Schneider and Bertrand Tavin.
\newblock Seasonal stochastic volatility and the {S}amuelson effect in
  agricultural futures markets, 2018.
\newblock arXiv:1802.01393.

\bibitem{schneider2015seasonal}
Lorenz Schneider and Bertrand Tavin.
\newblock Seasonal stochastic volatility and correlation together with the
  {S}amuelson effect in commodity futures markets, 2015.
\newblock arXiv:1506.05911.

\bibitem{schneider2014samuelson}
Lorenz Schneider and Bertrand Tavin.
\newblock From the {S}amuelson volatility effect to a {S}amuelson correlation
  effect: {E}vidence from crude oil calendar spread options, 2014.
\newblock arXiv:1401.7913.

\bibitem{LB2021}
Sergiy Ladokhin and Svetlana Borovkova.
\newblock Three-factor commodity forward curve model and its joint {P} and {Q}
  dynamics.
\newblock {\em Energy Economics}, 101:105418, 2021.

\bibitem{BBV2013}
Ole~E. Barndorff-Nielsen, Fred~Espen Benth, and Almut E.~D. Veraart.
\newblock Modelling energy spot prices by volatility modulated {L}évy-driven
  {V}olterra processes.
\newblock {\em Bernoulli}, 19(3):803--845, 2013.

\bibitem{LL2012}
Lingfei Li and Vadim Linetsky.
\newblock Time-changed {O}rnstein-{U}hlenbeck processes and their applications
  in commodity derivative models.
\newblock {\em Mathematical Finance}, 24(2):289--330, 2012.

\bibitem{Ogetbil2020}
Orcan \"Ogetbil.
\newblock {Extensions of Dupire Formula: Stochastic Interest Rates and
  Stochastic Local Volatility}, 2020.
\newblock arXiv:2005.05530.

\bibitem{jaeck2014samuelson}
Edouard Jaeck and Delphine Lautier.
\newblock Samuelson hypothesis and electricity derivative markets.
\newblock In {\em 31st International French Finance Association Conference,
  AFFI 2014}, page~24, 2014.

\bibitem{daalreexamining}
Elton Daal, Joseph Farhat, and Peihwang~P Wei.
\newblock Reexamining the maturity effect using extensive futures data.
\newblock {\em Department of Economics and Finance Working Papers, The
  University of New Orleans, 1991-2006. Paper 12.}, 2003.

\bibitem{OGH2022}
Orcan \"Ogetbil, Narayan Ganesan, and Bernhard Hientzsch.
\newblock Calibrating local volatility models with stochastic drift and
  diffusion.
\newblock {\em International Journal of Theoretical and Applied Finance},
  25(02):2250011, 2022.

\bibitem{CIR1985}
John~C. Cox, Jonathan~E. Ingersoll, and Stephen~A. Ross.
\newblock {A Theory of the Term Structure of Interest Rates}.
\newblock {\em Econometrica}, 53(2):385--407, 1985.

\end{thebibliography}

\end{document}